\documentclass[9pt,twocolumn,twoside]{osajnl}
\usepackage{threeparttable}
\usepackage{easyReview}
\journal{ao} % Choose journal (ao, aop, josaa, josab, ol)
\setboolean{shortarticle}{false} 
\ifthenelse{\boolean{shortarticle}}{\colorlet{color2}{color2b}}{\colorlet{color2}{color2}} % Automatically switches colors for short articles

\title{Absolute frequency measurement of molecular iodine hyperfine transitions at 647~nm}

\author[1]{Yao-Chin Huang}
\author[2]{Yu-Chan Guan}
\author[1]{Te-Hwei Suen}
\author[1,2]{Jow-Tsong Shy}
\author[1,3,*]{Li-Bang Wang}

\affil[1]{Department of Physics, National Tsing Hua University, Hsinchu 30013, Taiwan}
\affil[2]{Institute of Photonics Technologies, National Tsing Hua University, Hsinchu 30013, Taiwan}
\affil[3]{Frontier Research Center on Fundamental and Applied Sciences of Matters, National Tsing Hua University, Hsinchu 30013, Taiwan}

\affil[*]{Corresponding author: lbwang@phys.nthu.edu.tw}

\dates{Compiled \today}

\ociscodes{(300.6320) Spectroscopy, high-resolution; (300.6390) Spectroscopy, molecular; (300.6460) Spectroscopy, saturation; (140.2020) Diode lasers; (140.3425) Laser stabilization; (140.7300) Visible lasers.}

\doi{\url{http://dx.doi.org/10.1364/ao.XX.XXXXXX}}

\begin{abstract}
We report absolute frequency measurements of the molecular iodine P(46) 5-4 $a_{1}$, $a_{10}$, and $a_{15}$ hyperfine transitions at 647~nm with a fiber-based frequency comb. The light source is based on a Littrow-type external-cavity diode laser. A frequency stability of 5~$\times$~10$^{-12}$ at a 200~s integration time is achieved when the light source is stabilized to the P(46) 5-4 $a_{15}$ line. The pressure shift is determined to be -8.3(7)~kHz/Pa. Our determination of the line centers reached a precision of 21~kHz. The light source can serve as a reference laser for lithium spectroscopy ($2S \to 3P$).
\end{abstract}

\setboolean{displaycopyright}{false}

\begin{document}

\maketitle
\thispagestyle{fancy}

\ifthenelse{\boolean{shortarticle}}{\ifthenelse{\boolean{singlecolumn}}{\abscontentformatted}{\abscontent}}{}

\section{Introduction}
Molecular iodine ($I_2$) has played an important role in optical frequency standards for many applications. The most striking element of the iodine absorption spectrum is the long series of the $B-X$ system in the range from the green to near-IR and has very strong transition strength and narrow linewidth (<1~MHz) in the region close to the dissociation limit \cite{Cheng:02}. Molecular iodine is commonly used for frequency stabilization of laser. Stabilized laser by sub-Doppler saturation spectroscopy may be called a traditional method in molecular iodine spectroscopy. That good signal-to-noise ratio at the 532~nm of frequency doubled Nd:YAG laser gives excellent results \cite{Fang200676}. Stabilized laser as frequency reference can also be applied to fundamental metrology \cite{Huang:13,Hsiao:13}, atomic or molecular physics \cite{PhysRevLett.95.203001,huang2013precision,PhysRevA.82.042504,PhysRevLett.111.013002,PhysRevA.88.054501,Biesheuvel2016}, cavity QED system \cite{PhysRevA.92.013817} and optical clocks \cite{Daley2011,PhysRevA.91.023626,Kobayashi:16}.

Iodine absorption lines at 647~nm are very close to doubled wavelength (2$\times$323~nm) of the $2S \to 3P$ transition of atomic lithium (Li). Therefore, these iodine lines can be frequency references for Li atom research, e.g., the development of laser cooling \cite{PhysRevA.84.061406,PhysRevA.93.053424} and the measurement of hyperfine intervals \cite{PhysRevA.17.1394}. The frequency references at 647~nm using a simple iodine cell are particularly helpful in regards to studying lithium spectroscopy, especially for atomic physics laboratories where no optical frequency combs are available.

In this work, we perform Doppler-free saturation spectroscopy in an iodine vapor cell using modulation transfer spectroscopy. The effect of pressure shift is investigated by changing the cold finger temperature and the absolute transition frequency at zero pressure can be obtained. The saturated absorption signal is used to lock the laser frequency, and the absolute frequencies of the transitions are measured by an optical frequency comb. To our knowledge, there are no precision measurements around this wavelength, and our results provide useful input for the theoretical predictions.

\section{Experiment}
\begin{figure}[htbp]
\centering
\includegraphics[width=\linewidth]{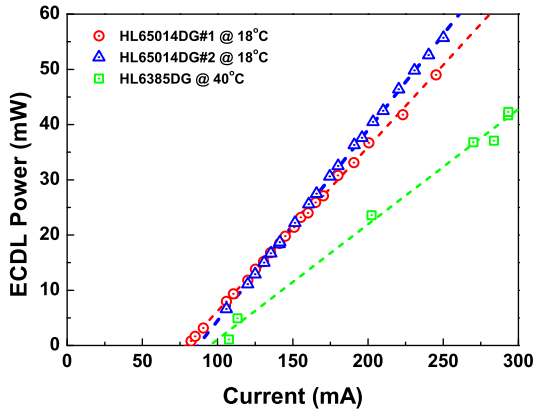}
\caption{The characteristics of light source at 647~nm consisted of a ECDL with two different diodes. The gain slop of blue line is 0.35~mW/mA.}
\label{ECDL_Power}
\end{figure}

The light source is a Littrow-type external-cavity diode laser (ECDL). We try two different laser diodes from Oclaro (model HL6385DG \& HL65014DG) in a standard $\phi$~5.6~mm package and show the characteristics in Fig. \ref{ECDL_Power} with Littrow configuration. 
The HL6385DG diode has a free-running wavelength of 642~nm at 25~$^{\circ}$C, so we increase its temperature to 40~$^{\circ}$C to shift the laser gain profile around 647~nm. However, the HL65014DG diode has a free-running wavelength  of 649~nm at room temperature and may be cooled to shift gain profile near 647~nm at 18~$^{\circ}$C. By grating feedback, the high-power single-mode semiconductor laser system can be operated at 647~nm. The characteristics of this diode laser system using HL65014DG are better than HL6385DG diode (e.g. threshold current, gain slope, and power). The laser is assembled in the Littrow configuration in order to establish a frequency selective external cavity. Mode-hope free tuning range of >5~GHz is achieved by controlling piezoelectric transducer and this allows us to observe the broad spectral features of interest for iodine molecular experiment.

\begin{figure}[htbp]
\centering
\includegraphics[width=1.0\linewidth]{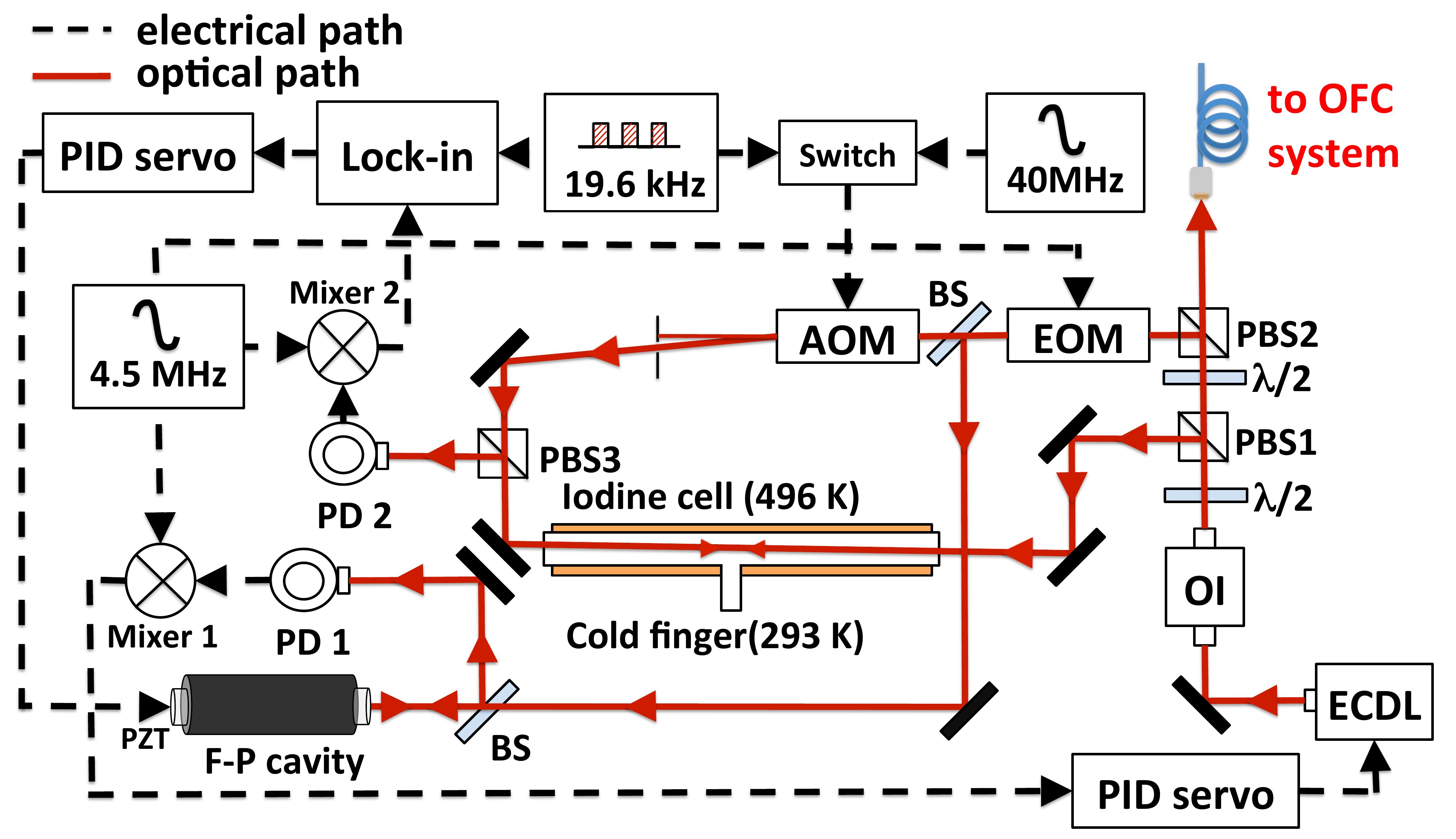}
\caption{Diagram of the experimental setup. ECDL, external-cavity diode laser; OI, optical isolator; $\lambda/2$, half-wave plate; PBS, polarizing beam splitter; BS, beam splitter; F-P cavity, Fabry-P$\rm \acute{e}$rot cavity; PD, photodiode; PI servo, electronic servo loop; EOM, electro-optic modulator; AOM, acousto-optic modulator; Lock-in, lock-in amplifier.}
\label{Fig1_Iodine_Setup}
\end{figure}

The experimental setup is shown in Fig. \ref{Fig1_Iodine_Setup}. The laser frequency is stabilized to a Fabry-P$\rm \acute{e}$rot cavity to reduce its linewidth by controlling the piezoelectric transducer voltage and driving current of the ECDL. As a result, the laser linewidth is reduced from approximately 2~MHz in free-running condition to 100~kHz when locked to the cavity. In general the power of light source is 35~mW after optical isolator. The ECDL output at 647~nm is used for our experiment to measure the iodine transition frequency by modulation transfer technique of saturation spectroscopy. One of the major advantages of this technique is that it provides a nearly flat baseline of an observed spectrum. A polarizing beam splitter (PBS1) and a half-wave plate are used to adequately distribute the power of the pump beam and probe beam. The optical frequency of the pump beam is phase modulated by an electro-optic modulator (EOM) at a modulation frequency of 4.5~MHz and shifted by an acousto-optic modulator (AOM) with a center driving RF frequency of 40~MHz. The RF frequency is chopped at 19.6~kHz by switching the AOM. The AOM is used to prevent interferometric baseline problems inside the iodine spectrometer. The pump and probe beams are overlapped at opposite directions in the iodine cell. The powers of the pump and probe beams are 15.2~mW and 1.6~mW, respectively. The diameter of both beams is approximately 2.5~mm. The probe beam is separated from the pump beam by a polarization beam splitter (PBS3) and detected by an amplified silicon photodiode. The signal is mixed in a double-balanced mixer with the EOM driving frequency to create the error signal. In order to obtain a background-free saturation signal, the mixer output is processed further by a lock-in amplifier (Stanford Research System SR830) with the chopping frequency of AOM as reference. The resulting signal is sent to a proportional, integral and derivative electronic servo circuit (Stanford Research System SIM960) to control the length of the Fabry-P$\rm \acute{e}$rot cavity so that the laser frequency is locked to the molecular transition.

A 58~cm long iodine vapor cell with plane windows fused to the cylindrical body at a slightly tilted angle with light path to avoid interference effects is used and the molecular iodine is contained in cold finger. The cold finger temperature is typically set at 20~$^{\circ}$C and is allowed to vary from 20~$^{\circ}$C to 30~$^{\circ}$C in order to study the effect of pressure shift. The stability of the temperature control of the cold finger is better than 10~mK. The iodine vapor pressure is related to the cold finger temperature of iodine cell by \cite {ja01302a050}
\begin{equation}
\rm log(\it P)=\rm - \frac{3512.830}{\it T}-2.013\times log(\it T)+ \rm18.37971,
\label{eq:iodine_pressure}
\end{equation}
where $\it P$ is the iodine vapor pressure in Pascals and $\it T$ is the cold finger temperature in Kelvins. The cell body is surrounded with multilayered glass fibre clothes and a tape heater, and its is maintained at 223~$^{\circ}$C. The heating of iodine vapor is necessary in order to populate the molecules to high-lying vibrational levels in the $B-X$ system.

\begin{figure}[bp]
\centering
\includegraphics[width=0.9\linewidth]{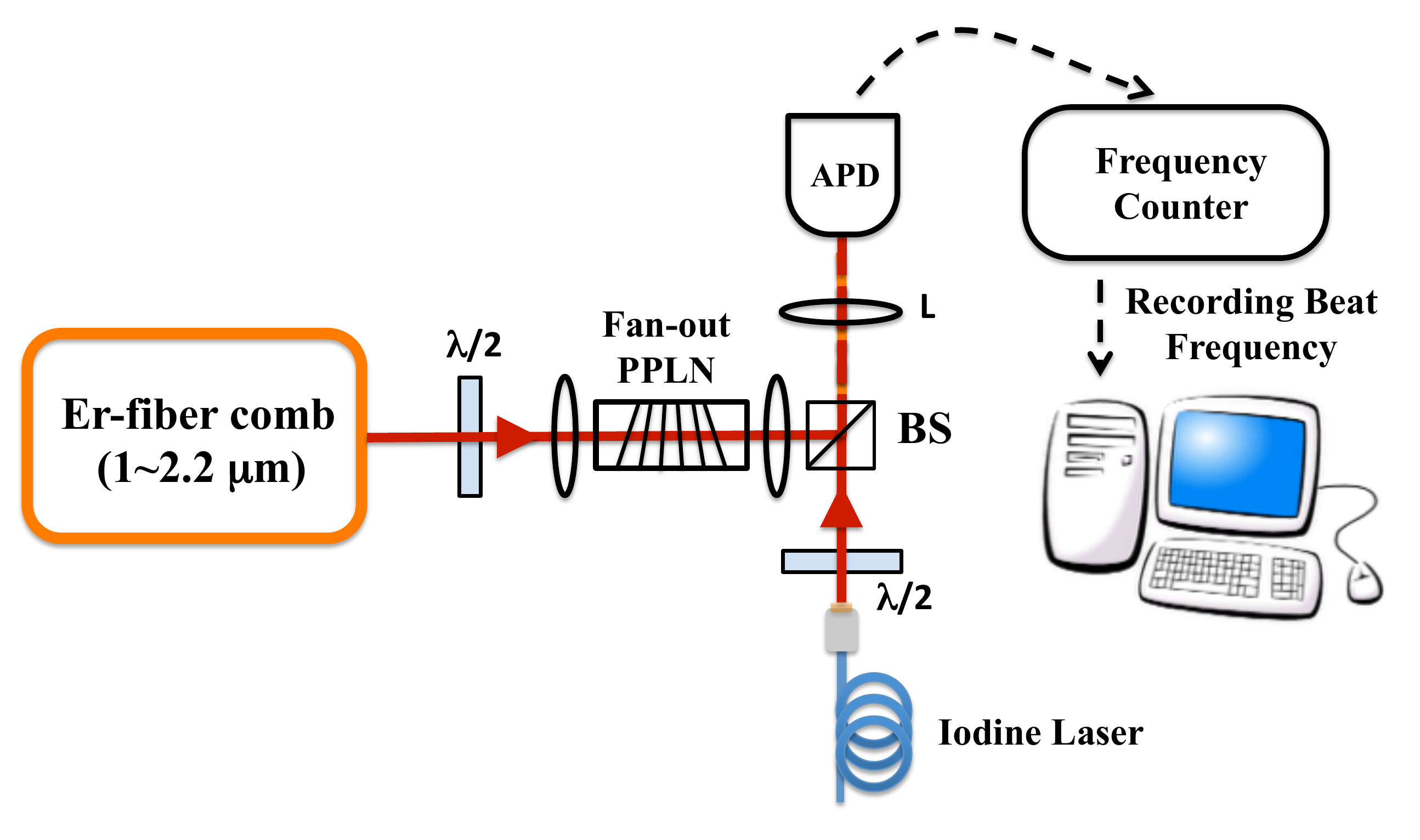}
\caption{Diagram of the absolute frequency measurement setup. $\lambda/2$, half-wave plate; PPLN, periodically poled lithium niobate; L, lens; BS, beam splitter; APD, amplified silicon photodiode.}
\label{Fig2_Measurement_Beat_Setup}
\end{figure}
A half-wave plate and a polarization beam splitter (PBS2, see Fig. \ref{Fig1_Iodine_Setup}) are used to separate the laser beam into two parts. A 5$\sim$7~mW portion of this beam is sent to an optical frequency comb with a single-mode fiber. When the laser is locked to the molecular iodine transition, the frequency of the 647~nm laser is measured by optical frequency comb. The frequency comb is based on a mode-locked erbium-doped fiber laser \cite{Peng2007,Inaba:06} pumped by a diode laser at 980~nm and operated at a repetition rate of approximately 250~MHz. The repetition rate frequency is phase locked to the Cs atomic clock (Hewlett Packard 5071A) by controlling the piezoelectric transducer mounted in cavity. The carrier envelop offset frequency of +21.4~MHz is detected using the $f-2f$ self-reference method and phase locked to microwave whose reference is connected to Cs atomic clock.
A heterodyne beat note is detected with amplified silicon photodiode between the 647~nm light and an fiber comb component, which is frequency doubled by passing the fundamental comb light through a fan-out type periodically poled lithium niobate (PPLN) nonlinear crystal with sum-frequency generation method  in Fig. \ref{Fig2_Measurement_Beat_Setup}. That is read with a frequency counter and is recorded by a computer. The accuracy of the fiber comb is achieved by referencing it to the Cs atomic clock. The Cs atomic clock is calibrated with hydrogen maser by National Time and Frequency Standard Laboratory of Chunghwa Telecom Co., Ltd. The precision of the fiber comb system has reached $9~\times~10^{-12}/\sqrt{\tau}$ for $\tau$ less than 1000~s (see Fig. \ref{Fig4_Allan_deviation}). Therefore, the fiber comb introduces negligible errors to our measurement. The iodine transition frequency $f_{\rm iodine}$ can be derived by the relation
\begin{eqnarray}
f_{\rm iodine}&=&\frac{f_{\rm pump}+f_{\rm probe}}{2}=f_{\rm laser}+\frac{f_{\rm AOM}}{2}\nonumber\\
&=&2\cdot f_{\rm comb}\pm f_{\rm beat}+\frac{f_{\rm AOM}}{2}\nonumber\\
&=&2(N\cdot f_{\rm rep}+f_{\rm ceo})\pm f_{\rm beat}+\frac{f_{\rm AOM}}{2},
\label{eq:absolute_frequency}
\end{eqnarray}
where $f_{\rm laser}$ is the 647~nm laser frequency, $f_{\rm rep}$ is the repetition rate frequency, $f_{\rm ceo}$ is the carrier envelop offset frequency, $f_{\rm beat}$ is the beat note frequency between the 647~nm laser and the fiber comb laser, and $f_{\rm AOM}$ is 40~MHz in our case. Here the mode number $N$ is a large integer to be determined as follow. We first use a wavelength meter to check our laser frequency to be within 200~MHz of the transition as predicted by the IodineSpec program \cite{Ispec} and iodine spectral atlas databook \cite{2000doppler}. 
\begin{figure}[t]
\centering
\includegraphics[width=1.0\linewidth]{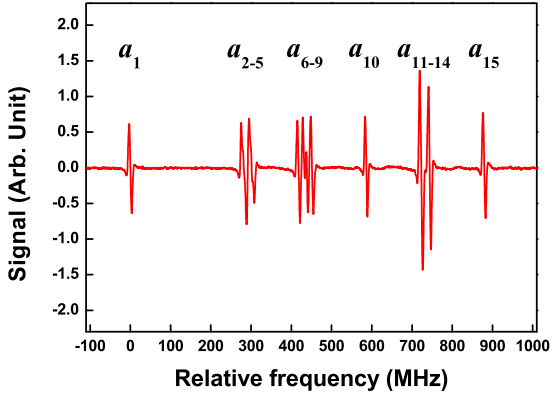}
\caption{Modulation transfer signals of the P(46) 5-4 transition. The signal-to-noise ratio (SNR) is 200 for $a_{15}$. A time constant of 30~ms and 12~dB/octave slop are used in obtaining the spectra.}
\label{Fig3_Longterm_Scan}
\end{figure}
We also change the repetition rate and the offset frequency of the comb laser on purpose to further confirm our measurements. 

\section{Results and Discussions}

\begin{figure}[tbp]
\centering
\includegraphics[width=1.0\linewidth]{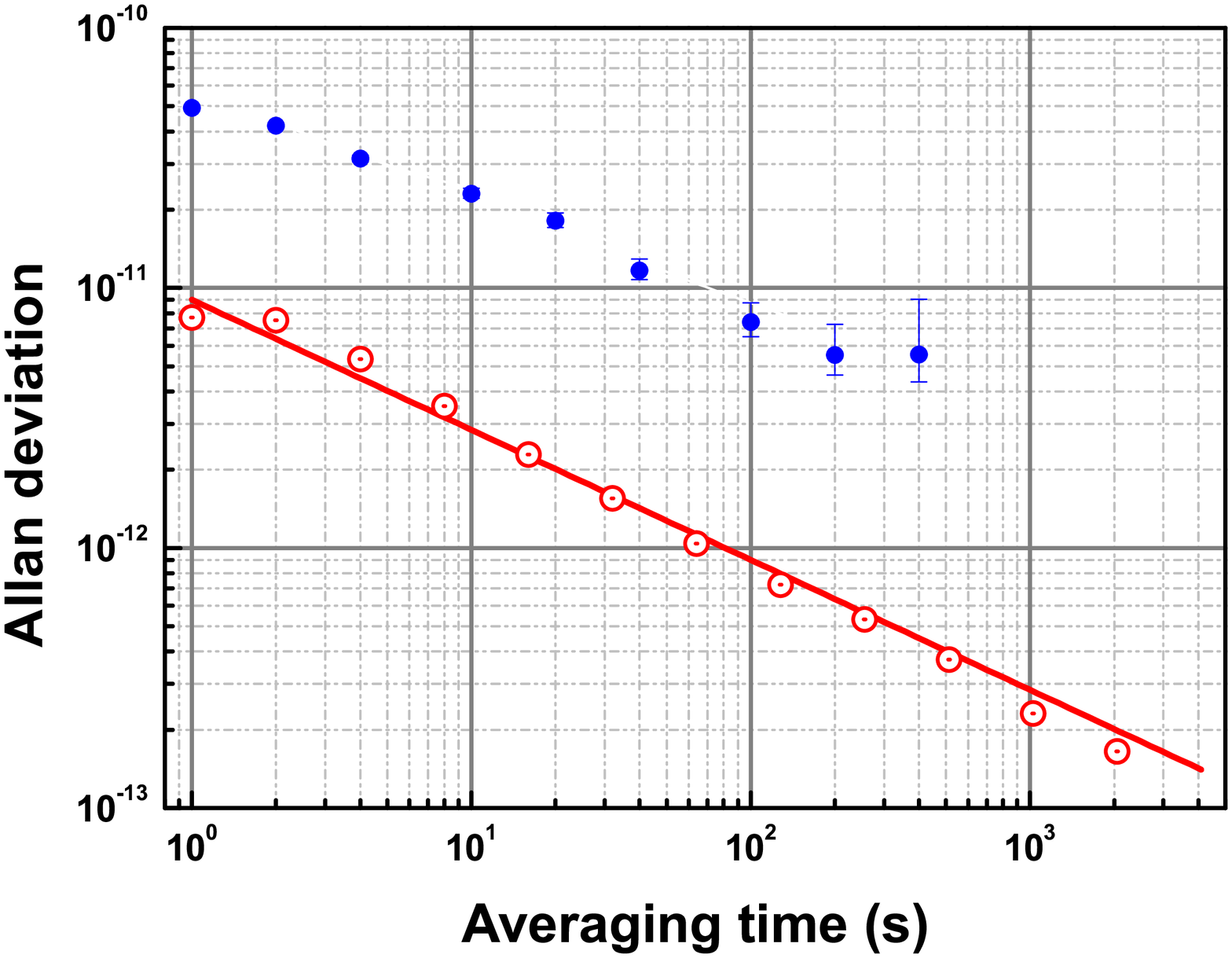}
\caption{(solid circle): Allan deviation $\sigma_{y}(\tau)$ is calculated from measuring beat frequency datas between the laser and frequency comb. The stability of the stabilized laser reaches a level of 5 $\times$ 10$^{-12}$ at a 200~s integration time. (open circle): Allan deviation of the Cs atomic clock \cite{clock}. (red line): $\sigma_{y}(\tau)~=~9\times10^{-12}/\sqrt{\tau}$. }
\label{Fig4_Allan_deviation}
\end{figure}

A continuous frequency scans across a full spectral range of the hyperfine structure could be realized without mode-hopping in Fig. \ref{Fig3_Longterm_Scan}. For the ground state of molecular iodine with an even rotational quantum ($J^{"}$ = 46, for our case), the rovibrational energy level is split into 15 sublevels. However, the three resolved lines $a_{1}$, $a_{10}$, and $a_{15}$ are clearly separated from other lines and are preferable for laser stabilization. We perform measurements on the $a_{1}$, $a_{10}$, and $a_{15}$ components of the P(46) 5-4 transition of $^{127}I_{2}$. The signal-to-noise ratio (SNR) of the $a_{15}$ component is approximately 200 for a lock-in time constant 30~ms and 12~dB/octave slop. To measure the transition frequency, partial light of laser is coupled into a single-mode fiber and sent to an optical frequency comb. The frequency stability of the laser locked to the P(46) 5-4 $a_{15}$ line is measured by recording the beat frequency between the laser and the frequency comb with a 100~ms gate time by a frequency counter (Pendulum CNT-90, $\Pi$ type). The total measurement time is 1700~s. The calculated Allan deviation is shown in  Fig. \ref{Fig4_Allan_deviation}. The stability of the stabilized laser reaches a level of 5~$\times$~10$^{-12}$ at a 200~s integration time.

\begin{figure}[htbp]
\centering
\includegraphics[width=1.0\linewidth]{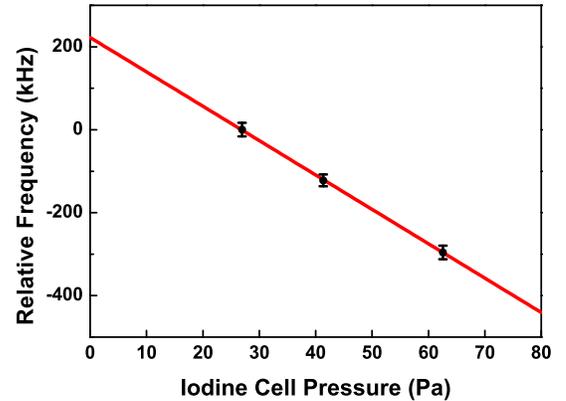}
\caption{Pressure shift of the $a_{15}$ transition frequency. The negative slope of 8.3(7)~kHz/Pa shows that the interaction due to collision is attractive. Each data point is obtained by a constant fit of 5 measurements.}
\label{Fig6_Pressure_Shift}
\end{figure}
In order to accurately determine the transition frequency at zero pressure, the pressure dependence must be carefully studied. 
We vary the cold finger temperature to change the iodine vapor pressure according to Eq. (\ref{eq:iodine_pressure}) and wait for minutes for temperature stabilization. Typical data are shown in Fig. \ref{Fig6_Pressure_Shift}. The result shows great linearity at the pressure range between 20 and 65~Pa. A linear fit is employed to extract the slope of the pressure shift. At our level of precision, we do not observe differences in the slopes among the three components we measure. Therefore, we combine all the results and obtain a coefficient of pressure shift as -8.3$\pm$0.7~kHz/Pa. We then use this coefficient to correct all the measured results to obtain the zero-pressure values. 

\begin{figure}[htbp]
\centering
\includegraphics[width=\linewidth]{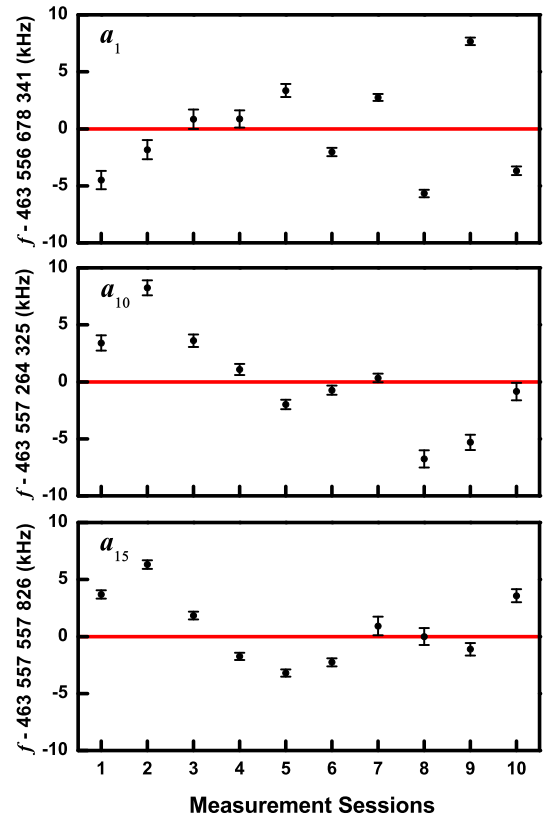}
\caption{Frequency measurements over 10 sessions at cold finger temperature of 20~$^{\circ}\rm C$. Each data point represents the mean value of 2000 measurements with gate time of 100~ms. The standard deviation of the measurements divided by square root of 2000 is assigned as the error bar of each point. The results for the three transitions $a_1$, $a_{10}$, and $a_{15}$ are 463~556~678~341(6)~kHz, 463~557~264~325(7)~kHz, and 463~557~557~826(6)~kHz, respectively.}
\label{Fig5_Iodine_a01-a10}
\end{figure}

Figure \ref{Fig5_Iodine_a01-a10} shows the results for the three components obtained over several days. Each data point consists of 2000 measurements with cold finger of 20~$^{\circ}\rm C$. During the 10 session points, we repeated the process of locking and unlocking the laser. The servo electronic offset, for example, a dc voltage offset between the baseline of the spectrum and the lock point, was adjusted by servo parameter to  <0.5~mV with average time to avoid the offset in the laser frequency. An estimate of the frequency stability ($\Delta f$) can be obtained by dividing the linewidth by the signal-to-noise ratio (SNR), leading to $\Delta f \approx$ 23~kHz. The sources of measurement uncertainties are summarized in Table \ref{tab:source-error}.
\section{Conclusion}
In summary, we have measured the absolute transition frequencies of the P(46) 5-4 $a_{1}$, $a_{10}$, and $a_{15}$ hyperfine transitions at 647~nm by modulation transfer saturation spectroscopy with the aid of an optical frequency comb. The results (see Table \ref{tab:Measured-Calculated}) are in agreement with the prediction of IodineSpec5, in which the expected uncertainty ($2\sigma$) in this wavelength is quoted to be $\pm$3~MHz \cite{Ispec}. The effect of pressure shift is important at our level of precision, and our measurement yields a linear dependence with a slop of -8.3(7)~kHz/Pa. Our precision measurements at 647~nm will help to reduce the uncertainties of the iodine atlas. In addition, our laser system can serve as a reference for $2S \to 3P$ transition of the lithium spectroscopy. 
\begin{table}[tbp]
\centering
\begin{threeparttable}
\caption{\bf Sources of Uncertainties~(kHz)}
\begin{tabular}{lrcc}
\hline
Source & &Correction & Error \\
\hline
Pressure shift && 224 & 19\\
Locking effect &&  & 5\\
Statistical stability &$a_1$:& --- &6\\
                             &$a_{10}$:& --- &7\\
                             &$a_{15}$:& --- &6\\
\hline
\end{tabular}
  \label{tab:source-error}
\end{threeparttable}
\end{table}

\begin{table}[htbp]
\centering
\begin{threeparttable}
\caption{\bf Results of the Transition Frequency and Hyperfine Splittings and Comparison to the Calculated Values~(kHz)}
\begin{tabular}{lccc}
\hline
P(46) 5-4 & Measured~\tnote{$a$} & Calculated & Deviation\\
\hline
$a_{1}$ & 463~556~678~565(21) & 463~556~679~461 & -896 \\
$a_{10}$ & 463~557~264~549(21) & 463~557~265~431 & -882 \\
$a_{15}$ & 463~557~558~050(21) & 463~557~558~984 & -934 \\
$a_{10}-a_{1}$& 585~984(30) & 585~970 & 14 \\
$a_{15}-a_{10}$& 293~501(30) & 293~553 & -52 \\
\hline
\end{tabular}
\begin{tablenotes}
\item[$a$]{The uncertainty is the combined error of the value in Fig. \ref{Fig5_Iodine_a01-a10} and the uncertainties listed in Table \ref{tab:source-error}.}
\end{tablenotes}
  \label{tab:Measured-Calculated}
\end{threeparttable}
\end{table}

\section*{Acknowledgment}
We gratefully acknowledge the financial support from the Ministry of Science and Technology
and the Ministry of Education of Taiwan. L.-B. Wang receives support from Kenda Foundation as a Golden-Jade fellow. We thank the National Time and Frequency Standard Laboratory of Chunghwa Telecom Co., Ltd. for providing the Cs atomic clock. We also thank Dr. H. Kn$\rm \ddot{o}$ckel for providing the IodineSpec5 software, which is essential for our experiment.

% Bibliography
\bibliography{sample}

\end{document}